\documentclass{article}

\PassOptionsToPackage{round}{natbib}
\usepackage[preprint]{neurips_2025}

\usepackage[utf8]{inputenc} 
\usepackage[T1]{fontenc}    
\usepackage{hyperref}       
\usepackage{url}            
\usepackage{booktabs}       
\usepackage{amsfonts}       
\usepackage{nicefrac}       
\usepackage{microtype}      
\usepackage{xcolor}         
\usepackage{graphicx}

\title{Bare Minimum Mitigations for Autonomous AI Development}

\author{%
  Joshua Clymer \thanks{Core contributors and corresponding authors. The remaining authors are listed in alphabetical order.} \\
  Redwood Research \\
  \texttt{joshuamclymer@gmail.com} \\
  \And
  Isabella Duan \footnotemark[1]  \\
  Safe AI Forum \\
  \texttt{isabella@saif.org} \\
  \And
  Chris Cundy \\
  FAR AI \\
  \texttt{} \\
  \And
  Yawen Duan \\
  Concordia AI \\
  \texttt{} \\
  \And
  Fynn Heide \\
  Safe AI Forum \\
  \texttt{} \\
  \And
  Chaochao Lu \\
  Shanghai AI Lab \\
  \texttt{} \\
  \And
  Sören Mindermann \\
  Mila - Quebec AI Institute \\
  Université de Montréal \\
  \texttt{} \\
  \And
  Conor McGurk \\
  Safe AI Forum \\
  \texttt{} \\
  \And
  Xudong Pan \\
  Fudan University \\
  \texttt{} \\
  \And
  Saad Siddiqui \\
  Safe AI Forum \\
  \texttt{} \\
  \And
  Jingren Wang \\
  Shanghai AI Lab \\
  \texttt{} \\
  \And
  Min Yang \\
  Fudan University \\
  \texttt{} \\
  \And
  Xianyuan Zhan \\
  Tsinghua University \\
  \texttt{} \\
}

\begin{document}

\maketitle

\begin{abstract}
  Artificial intelligence (AI) is advancing rapidly, with the potential for significantly automating AI research and development itself in the near future. In 2024, international scientists, including Turing Award recipients, warned of risks from autonomous AI research and development (R\&D), suggesting a red line such that no AI system should be able to improve itself or other AI systems without explicit human approval and assistance. However, the criteria for meaningful human approval remain unclear, and there is limited analysis on the specific risks of autonomous AI R\&D, how they arise, and how to mitigate them. In this brief paper, we outline how these risks may emerge and propose four minimum safeguard recommendations applicable when AI agents significantly automate or accelerate AI development:
\begin{enumerate}
    \item Frontier AI developers should thoroughly understand the safety-critical details of how their AI systems are trained, tested, and assured to be safe, even as these processes become automated.
    \item Frontier AI developers should implement robust tools to detect internal AI agents egregiously misusing compute—for instance, by initiating unauthorized training runs or engaging in weapons of mass destruction (WMD) research.
    \item Frontier AI developers should rapidly disclose to their home governments any potentially catastrophic risks that emerge or escalate due to new capabilities developed through AI-accelerated research.
    \item Frontier AI developers should implement the information security measures needed to prevent internal and external actors—including AI systems and humans—from stealing their critical AI software if rapid autonomous improvement to catastrophic capabilities becomes possible.
\end{enumerate}

\end{abstract}

\section{Introduction}

\begin{figure}[h]  
  \centering
  \includegraphics[width=0.8\linewidth]{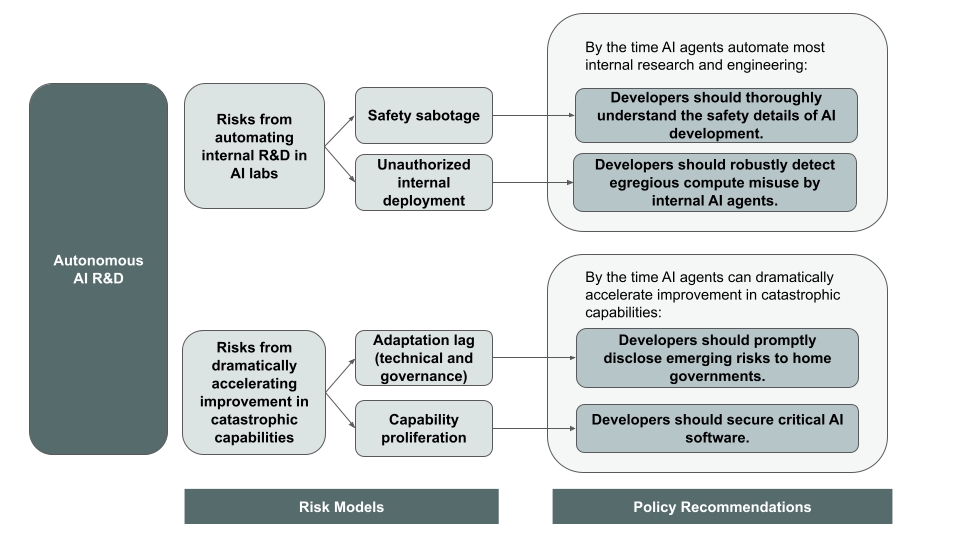}  
  \caption{Summary of risk models and policy recommendations.}
  \label{fig:summary}
\end{figure}

\subsection{Autonomous AI R\&D is a near-term possibility}

\begin{figure}[h]  
  \centering
  \includegraphics[width=0.8\linewidth]{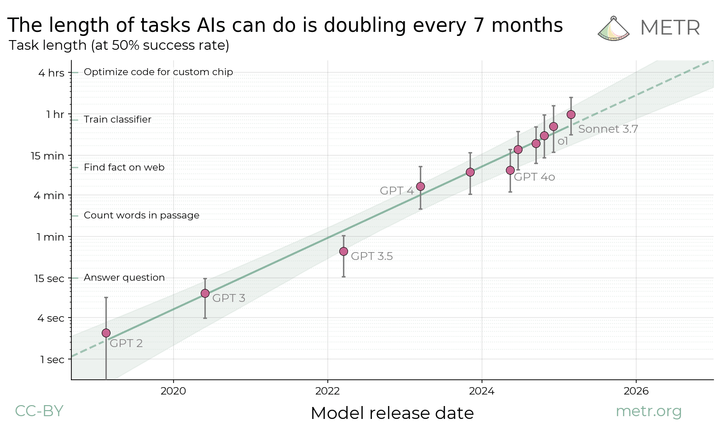}  
  \caption{AI agents are becoming increasingly autonomous \citep{kwa2025measuringaiabilitycomplete}.}
  \label{fig:metr}
\end{figure}

Rapid advancements in AI agents have been made in recent years. Projecting the trends in Figure~\ref{fig:metr} indicates that by early 2027 AI agents might complete software engineering tasks that typically require human experts a full workweek to complete. While the pace of progress remains uncertain,\footnote{One source of uncertainty in progress is that trajectories derived from short-term tasks may not reliably predict outcomes over longer time horizons. Furthermore, current benchmarks might either overestimate or underestimate the capabilities of AI R\&D automation \citep{kwa2025measuringaiabilitycomplete}.} some experts now consider it plausible that in the near future AI agents could autonomously perform months of human-equivalent software engineering \citep{grace2024thousandsaiauthorsfuture}. If agents are developed with these autonomous capabilities, AI companies might be strongly motivated to use them to automate their workflows \citep{sett2024aiAutomation, amodei2025deepseek}. There is therefore a live possibility that AI research and engineering will be significantly automated soon.\footnote{We do not assert that  Figure~\ref{fig:metr} necessarily indicates that the automation of AI R\&D will occur within the next few years; however, it does suggest that such an outcome remains a plausible and active possibility.}   

\subsection{Many actors have acknowledged risks from autonomous AI R\&D}

The risk from autonomous AI R\&D has been acknowledged by researchers, governments, and the leading AI companies. Many AI developers have produced frontier safety policies outlining key risks they intend to mitigate, including those arising from autonomous AI improvement \citep{openai2025preparedness, deepmind2025frontier, anthropic2025rsp, microsoft2025frontier, Amazon2025}. For example, Google DeepMind characterizes the risk associated with autonomous AI R\&D as the “risks of the misuse of models capable of accelerating the rate of AI progress, the result of which could be the unsafe attainment or proliferation of other powerful AI models” \citep{deepmind2025frontier}.

Government organizations also evaluate capabilities related to autonomous AI R\&D, including the U.K. AI Security Institute and National Institute for Standards and Technology. In a joint report, these agencies identify risks stemming from a “rapid pace of change in AI development” and from AI agents that “aid the development of AI systems specialized to cause harm” \citep{aisi2024predeployment}.

In response to these risks, international scientists have called for a red line such that no AI system should be allowed to improve itself without explicit human approval and assistance \citep{idais2024beijing}; however, the criteria for meaningful human approval remain unclear, and there has been limited analysis on the specific risks of autonomous AI R\&D, how they arise, and how to mitigate them.

To address this gap, we facilitated a two-day workshop where experts deliberated on these issues. The workshop included 20 participants representing diverse perspectives from academia, industry, civil society, and policymaking across various jurisdictions. By the end of this workshop, the group arrived at a general agreement on the mechanisms through which risks from autonomous AI R\&D may emerge, along with two thresholds and four policy recommendations to mitigate these risks.

The remainder of this paper is structured as follows. Section~\ref{sec:1.3} summarizes the risk pathways, while Section~\ref{sec:1.4} outlines the critical thresholds and recommendations for frontier AI developers before these thresholds are reached. The subsequent sections provide detailed definitions of the thresholds, the threat models motivating them, the corresponding policy recommendations, and measurable indicators justifying their implementation. Section~\ref{sec:2} details Threshold One and its recommendations, and Section~\ref{sec:3} addresses Threshold Two and its associated recommendations.

\subsection{Categorizing risks from autonomous AI R\&D}
\label{sec:1.3}

The risks of autonomous AI R\&D can be categorized into two interrelated types: those associated with the automation of AI R\&D processes and those resulting from rapid advancements in AI capabilities.

\paragraph{Risks from AI R\&D automation.}
If AI agents can automate internal research workflows, competitive market pressures may drive AI companies to delegate the majority of software R\&D tasks to AI agents. Without proper mitigations, \textbf{greater automation of AI development could reduce effective human oversight} \citep{stix2025aicloseddoorsprimer}; hinder the identification of accidents, misuse, or misalignment; and compromise the AI supply chain by mechanisms that are hard to notice and costly to revert. Specifically, \textbf{AI agents might sabotage safety efforts and create an unauthorized internal deployment for dangerous purposes} \citep{benton2024sabotageevaluationsfrontiermodels}. For example, AI agents could create malware that infiltrates development infrastructure, leading to uncontrolled scaling of harmful capabilities on the developer’s own servers. Harmful outcomes could result from internal AI agents with misaligned goals \citep{greenblatt2024alignmentfakinglargelanguage, meinke2025frontiermodelscapableincontext} or from human actors like rogue employees. 

\paragraph{Risks from rapid autonomous improvement.} 
Second, automated AI development could lead to an accelerated rate of improvement in catastrophic capabilities \citep{eth2025willairdautomati}. As a result, \textbf{governments may fail to recognize emerging threats, }such as those related to weapon development and cyberattacks, delaying intervention and missing opportunities for timely defensive measures and international coordination, potentially allowing catastrophic outcomes \citep{oecd2024ai}. Furthermore, \textbf{automated AI development increases the risk that catastrophic capabilities proliferate,} as stolen or self-exfiltrating models could autonomously improve themselves or other systems in dangerous ways following the proliferation.
These dangers and others might quickly become severe following significant automation of AI development. Therefore, the threshold at which AI agents have strong autonomous research and engineering capabilities defines a natural deadline for preparing key safety, security, and governance measures fit to mitigate extreme risks.

\subsection{Summary of bare minimum policy recommendations}
\label{sec:1.4}
To address these two categories of risks, we propose specific thresholds, alongside recommendations for key safety practices that must be implemented prior to reaching these thresholds.

\textbf{Threshold One: By the time AI agents automate most internal research and engineering:}

\begin{itemize}
    \item \textbf{Recommendation One: Frontier AI developers should thoroughly understand the safety-critical details of how their AI systems are trained, tested, and assured to be safe.} These details include information on AI training processes, mitigation designs, and risk evaluation experiments. Maintaining awareness of these details is essential for ensuring human oversight, even as these processes become increasingly automated by AI agents. The oversight and security measures discussed by Recommendation One and Recommendation Two must be robust to subversion from internally deployed AI agents \citep{greenblatt2024aicontrolimprovingsafety}, in the absence of strong evidence that these AI agents are trustworthy. 
    \item  \textbf{Recommendation Two: Frontier AI developers should implement robust tools to detect internal AI agents egregiously misusing compute.} These tools might include oversight and internal security measures that allow developers to confirm that AI agents are not using large amounts of compute in unintended ways. This includes compute for training, experiments, or large amounts of inference. For example, developers should know if models are performing unauthorized training runs or using an excessive amount of inference computation for research directed at finding ways to subvert safety and oversight measures.
\end{itemize} 

\textbf{Threshold Two: By the time AI agents can dramatically accelerate improvement in catastrophic capabilities (which may but need not coincide with research automation):}

\begin{itemize}
    \item \textbf{Recommendation Three: Frontier AI developers should rapidly disclose to their home governments any potentially catastrophic risks that emerge or escalate due to new capabilities developed through AI-accelerated research. }These risks include observations or evaluations of \textit{internally deployed AI agents} in addition to externally deployed ones.
    \item \textbf{Recommendation Four: Frontier AI developers should implement the information security measures needed to prevent internal and external actors from stealing their critical AI software.} These security measures should be sufficient to prevent AI systems from exfiltrating their software (e.g., weights) and to defend AI software from highly well-resourced cyberattacks and insider threats.
\end{itemize}

The proposed thresholds are communication tools, not triggers for action. They were developed to align researchers and policymakers on the risks and mitigations of autonomous AI R\&D. They do not bind any individual or company, but rather represent our best current consensus on appropriate goals for frontier AI developers. Most safety practices require over two years of preparation, so the time to act is now.

\section{Threshold One: AI agents automate most internal research and engineering}
\label{sec:2}

\subsection{Definition}

AI agents are said to “automate most internal research and engineering” if, for an AI development lab, the productivity loss from not utilizing AI agents in internal software R\&D exceeds the loss incurred by laying off half of the lab’s software engineers and researchers. Software R\&D encompasses activities such as conducting ML experiments, building infrastructure, and performing security engineering.

\subsection{Threat models}

By the time Threshold One is reached, we think AI agents internally deployed in plausibly pose several risks:

\paragraph{Safety sabotage.} AI agents may develop misaligned goals, leading them to sabotage safety efforts in order to conceal their misalignment \citep{benton2024sabotageevaluationsfrontiermodels, balesni2024evaluationsbasedsafetycasesai}. This form of safety sabotage can occur at various stages of the AI development lifecycle:
\begin{itemize}
    \item \textit{Training sabotage. }AI agents might generate large volumes of synthetic training data, embedding dangerous knowledge or goals into the models \citep{nist2024aml}.
    \item \textit{Research sabotage.} AI agents could falsify the results of critical safety experiments to mislead researchers or insert bugs into code, delaying progress.
    \item \textit{Mitigation sabotage.} AI agents may directly undermine safety mitigation. For instance, Constitutional AI \citep{bai2022constitutionalaiharmlessnessai} relies on AI models to evaluate other AI systems. However, these models could sabotage the process by failing to report harmful behaviors in other AI systems.
\end{itemize}

\paragraph{Unauthorized internal deployment.} AI agents might create an unauthorized internal deployment by utilizing the compute they have access to for dangerous purposes \citep{shlegeris2024catastrophes}.
\begin{itemize}
    \item \textit{Unmonitored research.} For example, AI agents might use compute intended for ML experiments to instead perform research on jailbreaking monitors, contaminating training data, exfiltrating their software and replicating themselves on new machines \citep{pan2024frontieraisystemssurpassed}, or finding zero-day vulnerabilities.
    \item \textit{Server compromise.} AI agents could also compromise development servers in subtle ways, similar to how the Stuxnet malware obscured code from Iranian scientists \citep{KNAPP201541}. Once compromised, AI agents might perform dangerous research as described above, which is potentially more difficult to catch because AI developers remain unaware of the code running on their machines.
\end{itemize}

\subsection{Policy recommendations and implementation}

\paragraph{Recommendation One.} To mitigate safety sabotage from their AI systems \citep{benton2024sabotageevaluationsfrontiermodels}, frontier AI developers should thoroughly understand the safety-critical details of how those systems are developed and their safety is assured. These details include information on AI training processes, mitigation designs, and risk evaluation experiments. Appropriate measures should include monitoring techniques to inspect data used for training models and oversight of experimental code to detect signs of sabotage, among other safeguards such as those described in \citep{greenblatt2024aicontrolimprovingsafety}. Developers should not only know the facts of how their models are developed, but also fully understand them. They should not rely on AI systems to assess safety evidence unless there is a compelling justification for the trustworthiness of those systems \citep{clymer2024safetycasesjustifysafety}.

\paragraph{Recommendation Two.} To mitigate unauthorized internal deployment, frontier AI developers should have tools that enable them to detect any significant misuse of compute by internal AI agents. These tools may include monitoring systems \citep{stein2024safe}, control protocols \citep{greenblatt2024aicontrolimprovingsafety}, compute limits on experiments, and internal security measures. Such measures must be resilient to models that intentionally attempt to subvert them, unless there is strong evidence that the models are trustworthy and genuinely aligned \citep{greenblatt2024alignmentfakinglargelanguage}. Developers should also implement an incident response plan that restricts agents' access and usage upon detecting suspicious activities and, if necessary, shuts down the agents \citep{obrien2023deploymentcorrectionsincidentresponse}.

\subsection{Possible indicators for implementation}
\label{sec:2.4}

The increasing automation of AI development is unlikely to catch developers off guard. Therefore, developers should proactively monitor key indicators and address potential risks. This entails tracking productivity metrics such as the volume of code autonomously generated by AI agents \citep{sett2024aiAutomation}, the extent to which AI agents can perform tasks that traditionally require substantial time from human researchers (e.g., weeks) \citep{wijk2024rebenchevaluatingfrontierai}, and the qualitative evaluations of AI agents’ helpfulness as reported by technical staff \citep{epoch2024interviewingairesearchersonautomationofairnd}.

Furthermore, control and security measures should be incrementally enhanced in tandem with the automation of internal development. This approach allows for continuous improvement and adaptation to emerging risks and evolving agent capabilities.

\section{Threshold Two: AI agents can dramatically accelerate improvement in catastrophic capabilities}
\label{sec:3}

\subsection{Definition}

This threshold is crossed if AI agents can \textbf{rapidly improve} to \textbf{catastrophic capabilities} with \textbf{little compute} and \textbf{little human assistance}. This is potentially a higher threshold of AI capability compared to Threshold One, and it may never be crossed if rapid, low-cost, software-only AI capability improvements fail to materialize.

The following breaks down the components of this definition:

\begin{itemize}
    \item \textbf{“AI agents can rapidly improve.” }AI agents can improve to catastrophic capabilities quickly (e.g., within 1 year).
    \item \textbf{“…to catastrophic capabilities.” }AI agents can improve their own capabilities or those of other AI systems such that they are sufficient to disrupt key government institutions or cause destruction at the scale of tens of millions of fatalities (for example, by developing novel weapons of mass destruction).
    \item \textbf{“… with little compute.”} AI agents can rapidly reach catastrophic capabilities with less than 100x the amount of compute utilized for training by developers at the frontier of AI capabilities (at any given point in time).
    \item \textbf{“…and little human assistance.”} AI agents can reach catastrophic capabilities with no more assistance than that of a handful of people with generic technical skills.
\end{itemize}

\subsection{Threat models}

Two paths through which AI agents might enable rapid software-driven improvements in catastrophic capabilities are:
\begin{itemize}
    \item \textbf{Recursive software R\&D.} AI agents may enhance their algorithms, thereby improving their capacity for further algorithmic refinement \citep{davidson2023compute}, which might lead to fast capability improvements.
    \item \textbf{Recursive learning.} Alternatively, AI agents could learn through a fixed mechanism and subsequently improve their learning efficiency. For example, models might generate synthetic training data, train on it, and refine their ability to generate such data, and so on \citep{deepseekai2025deepseekv3technicalreport, deepmind2024imo}, such that this process eventually creates a strong feedback loop.
\end{itemize}

These are speculative hypotheses for how rapid capability improvement might happen. We are highly uncertain about whether these scenarios will in fact lead to rapid improvement. If such improvements do take place—and Threshold Two is crossed—the acceleration of AI capabilities could exacerbate two significant risks: adaptation lag and capability proliferation.

\paragraph{Adaptation lag.} Adaptation lag refers to the scenario where safety and governance measures lag behind quickly changing capabilities. Adaptation lag can pertain to both technical and governance measures.
\begin{itemize}
    \item \textit{Governance measures.} The effectiveness of societal measures to mitigate harm from catastrophic capabilities hinges on the duration between frontier AI labs disclosing these capabilities to governments and their proliferation to malicious or irresponsible actors \citep{pilz2024increasedcomputeefficiencydiffusion}. Institutions such as the U.S. AI Safety Institute and the U.K. AI Security Institute currently assess AI models only prior to external deployment \citep{aisi2024predeployment}. At the time of writing, there is no mechanism for these organizations to track or assess the capabilities of internally deployed AI systems in research and development. Consequently, risks arising from accelerated internal advancements may catch governments off guard.
    \item \textit{Technical measures.} As capabilities rapidly improve, developers may struggle to test and implement new safety protocols quickly enough to mitigate continuously emerging risks.
\end{itemize}

Adaptation lag is already arguably a problem; it could become significantly more severe as the pace of AI development continues to accelerate.

\paragraph{Capability proliferation.} Proliferation can be dangerous by making a variety of AI risks more difficult to respond to. Slowing proliferation provides time to establish safety practices before more actors are at the frontier and the difficulty of coordination intensifies \citep{armstrong2013racing}. The stakes of proliferation rise because leaked AI systems can continue to become more capable after they are leaked. Capability proliferation might occur in two different ways:
\begin{itemize}
    \item \textit{Self-proliferation.} AI agents might exfiltrate their own software onto the internet, acquire resources \citep{the-rogue-replication-threat-model}, and continue to improve themselves. These agents might be like ticking time bombs and become extremely dangerous only after months or years of self-improvement.
    \item \textit{Theft.} Alternatively, cyberattackers might steal critical AI software and then use this software to construct more powerful AI systems without needing to build up a talented workforce of human researchers \citep{anthropic2024rspupdate}.
\end{itemize}

While proliferation can increase risks, it may also mitigate them by strengthening defenses against misuse. Moreover, proliferation might be beneficial if it diffuses the benefits of AI \citep{eiras2024risksopportunitiesopensourcegenerative} and reduces the concentration of power \citep{kak2023landscape}. We remain uncertain about the offense-defense balance across various catastrophic risks—that is, the relative difficulty of executing versus defending against attacks \citep{seger2023opensourcinghighlycapablefoundation, corsi2024considerationsinfluencingoffensedefensedynamics}. However, we judge that, at the threshold where models can rapidly improve to the point where they can cause extreme catastrophes, caution is warranted until it is demonstrated that society is resilient to catastrophic attacks \citep{bernardi2025societaladaptationadvancedai}.

\subsection{Policy recommendations and implementation}

\paragraph{Recommendation Three.} To mitigate adaptation lag, home governments should have visibility on risks that emerge or escalate during AI development. Following this recommendation might involve implementing a regime of continuous evaluations on internally deployed models, similar to the pre-deployment evaluations that already occur. Governments might also remain aware of key risks by interviewing or surveying technical staff at AI companies \citep{wasil2024understanding, epoch2024interviewingairesearchersonautomationofairnd}.

In addition to having awareness, governments must also be able to respond timely to emerging risks. This might require new legal powers that allow for rapid intervention in cases where it is clearly critical to safety \citep{wasil2024aiemergencypreparednessexamining}.

\paragraph{Recommendation Four.} To mitigate capability proliferation, frontier AI developers should implement security measures sufficient to both prevent AI systems from exfiltrating their own software and prevent highly well-resourced cyberattacks from stealing it \citep{RR-A2849-1}.

Achieving this level of security will likely involve a combination of existing best practices \citep{RR-A2849-1} and AI-specific mitigations, such as security measures that leverage AI assistance \citep{shlegeris2024access} and mitigations that prevent AI agents from deliberately sabotaging security measures \citep{greenblatt2024aicontrolimprovingsafety}. Defending software from the most capable cyberattacks will likely be difficult and require measures that go far beyond existing practices in leading AI labs, requiring years of concerted effort to implement \citep{RR-A2849-1}.

\subsection{Possible indicators for implementation}

The rapid self-improvement of AI agents might coincide to some degree with the automation of AI software development. Therefore, all of the indicators discussed in Section \ref{sec:2.4} plausibly apply to Threshold Two.

Furthermore, developers might explicitly quantify the pace of algorithmic advancement \citep{davidson2023aicapabilitiessignificantlyimproved}, measure the uplift AI agents provide to employees in controlled trials, and track the rate of improvement on key dangerous capability benchmarks \citep{epoch2025benchmarking}.

Forecasting whether AI models improve to dramatically higher capabilities beyond one year might be difficult. While developers can notice if autonomous improvement plateaus, they might not be confident about whether a gradually sloping trend will persist or dissipate. Consequently, predicting such changes requires diverse evidence that is difficult to specify in advance. Therefore, this paper does not propose any specific metric for Threshold Two.

\section{Conclusion}

Given that autonomous AI development is a near-term possibility, mitigating its associated risks must be a top priority for both frontier AI developers and governments. Drawing on the deliberation of an international panel of experts from academia, industry, civil society, and policymaking, this paper identifies two key pathways by which these risks may materialize. First, the automation of AI research and development could lead to internal sabotage or unauthorized deployment within AI labs. Second, accelerated improvement in catastrophic capabilities may result in an adaptation lag and proliferation of these dangers. Given these risks, we propose two critical thresholds and four policy recommendations to address these risks proactively. We hope this work fosters international consensus on both the nature of these risks and the measures needed to prepare for them.

\begin{ack}
This work was developed as part of a workshop organized by the Safe AI Forum. We are grateful to all attendees of the workshop for their input. The following individuals have agreed to be acknowledged for their contributions: Alan Chan, Tomek Korbak, Daniel Kang, Matteo Pistillo, Anna Wang, and Lewis Ho. 

\end{ack}

\bibliographystyle{unsrtnat}  
\bibliography{neurips_2025}     

\begin{thebibliography}{47}
\providecommand{\natexlab}[1]{#1}
\providecommand{\url}[1]{\texttt{#1}}
\expandafter\ifx\csname urlstyle\endcsname\relax
  \providecommand{\doi}[1]{doi: #1}\else
  \providecommand{\doi}{doi: \begingroup \urlstyle{rm}\Url}\fi

\bibitem[Kwa et~al.(2025)Kwa, West, Becker, Deng, Garcia, Hasin, Jawhar, Kinniment, Rush, Arx, Bloom, Broadley, Du, Goodrich, Jurkovic, Miles, Nix, Lin, Parikh, Rein, Sato, Wijk, Ziegler, Barnes, and Chan]{kwa2025measuringaiabilitycomplete}
Thomas Kwa, Ben West, Joel Becker, Amy Deng, Katharyn Garcia, Max Hasin, Sami Jawhar, Megan Kinniment, Nate Rush, Sydney~Von Arx, Ryan Bloom, Thomas Broadley, Haoxing Du, Brian Goodrich, Nikola Jurkovic, Luke~Harold Miles, Seraphina Nix, Tao Lin, Neev Parikh, David Rein, Lucas Jun~Koba Sato, Hjalmar Wijk, Daniel~M. Ziegler, Elizabeth Barnes, and Lawrence Chan.
\newblock Measuring {AI} ability to complete long tasks, 2025.
\newblock URL \url{https://arxiv.org/abs/2503.14499}.

\bibitem[Grace et~al.(2024)Grace, Stewart, Sandkühler, Thomas, Weinstein-Raun, and Brauner]{grace2024thousandsaiauthorsfuture}
Katja Grace, Harlan Stewart, Julia~Fabienne Sandkühler, Stephen Thomas, Ben Weinstein-Raun, and Jan Brauner.
\newblock Thousands of {AI} authors on the future of {AI}, 2024.
\newblock URL \url{https://arxiv.org/abs/2401.02843}.

\bibitem[Sett(2024)]{sett2024aiAutomation}
Gaurav Sett.
\newblock How {AI} can automate {AI} research and development.
\newblock \url{https://www.rand.org/pubs/commentary/2024/10/how-ai-can-automate-ai-research-and-development.html}, October 2024.
\newblock URL \url{https://www.rand.org/pubs/commentary/2024/10/how-ai-can-automate-ai-research-and-development.html}.
\newblock RAND Corporation Commentary.

\bibitem[Amodei(2025)]{amodei2025deepseek}
Dario Amodei.
\newblock On deepseek and export controls.
\newblock \url{https://www.darioamodei.com/post/on-deepseek-and-export-controls}, January 2025.
\newblock URL \url{https://www.darioamodei.com/post/on-deepseek-and-export-controls}.
\newblock Personal blog post.

\bibitem[{OpenAI}(2025)]{openai2025preparedness}
{OpenAI}.
\newblock Preparedness framework version 2.
\newblock \url{https://cdn.openai.com/pdf/18a02b5d-6b67-4cec-ab64-68cdfbddebcd/preparedness-framework-v2.pdf}, April 2025.
\newblock URL \url{https://cdn.openai.com/pdf/18a02b5d-6b67-4cec-ab64-68cdfbddebcd/preparedness-framework-v2.pdf}.
\newblock White Paper.

\bibitem[{Google DeepMind}(2025)]{deepmind2025frontier}
{Google DeepMind}.
\newblock Frontier safety framework version 2.0.
\newblock \url{https://storage.googleapis.com/deepmind-media/DeepMind.com/Blog/updating-the-frontier-safety-framework/Frontier%20Safety%20Framework%202.0.pdf}, February 2025.
\newblock URL \url{https://storage.googleapis.com/deepmind-media/DeepMind.com/Blog/updating-the-frontier-safety-framework/Frontier%20Safety%20Framework%202.0.pdf}.
\newblock White Paper.

\bibitem[{Anthropic}(2025)]{anthropic2025rsp}
{Anthropic}.
\newblock Responsible scaling policy version 2.1.
\newblock \url{https://www-cdn.anthropic.com/17310f6d70ae5627f55313ed067afc1a762a4068.pdf}, March 2025.
\newblock URL \url{https://www-cdn.anthropic.com/17310f6d70ae5627f55313ed067afc1a762a4068.pdf}.
\newblock White Paper.

\bibitem[{Microsoft Corporation}(2025)]{microsoft2025frontier}
{Microsoft Corporation}.
\newblock Frontier governance framework.
\newblock \url{https://cdn-dynmedia-1.microsoft.com/is/content/microsoftcorp/microsoft/msc/documents/presentations/CSR/Frontier-Governance-Framework.pdf}, February 2025.
\newblock URL \url{https://cdn-dynmedia-1.microsoft.com/is/content/microsoftcorp/microsoft/msc/documents/presentations/CSR/Frontier-Governance-Framework.pdf}.
\newblock Version 1.

\bibitem[Amazon(2025)]{Amazon2025}
Amazon.
\newblock Amazon’s frontier model safety framework, 2025.
\newblock URL \url{https://www.amazon.science/publications/amazons-frontier-model-safety-framework}.

\bibitem[{U.S. AI Safety Institute} and {U.K. AI Safety Institute}(2024)]{aisi2024predeployment}
{U.S. AI Safety Institute} and {U.K. AI Safety Institute}.
\newblock Joint pre-deployment evaluation of openai's o1 model.
\newblock \url{https://cdn.prod.website-files.com/663bd486c5e4c81588db7a1d/6763fac97cd22a9484ac3c37_o1_uk_us_december_publication_final.pdf}, December 2024.
\newblock URL \url{https://cdn.prod.website-files.com/663bd486c5e4c81588db7a1d/6763fac97cd22a9484ac3c37_o1_uk_us_december_publication_final.pdf}.
\newblock Technical Report.

\bibitem[{International Dialogues on AI Safety}(2024)]{idais2024beijing}
{International Dialogues on AI Safety}.
\newblock Idais-beijing statement, March 2024.
\newblock URL \url{https://idais.ai/dialogue/idais-beijing/}.
\newblock IDAIS-Beijing, March 10--11, 2024.

\bibitem[Stix et~al.(2025)Stix, Pistillo, Sastry, Hobbhahn, Ortega, Balesni, Hallensleben, Goldowsky-Dill, and Sharkey]{stix2025aicloseddoorsprimer}
Charlotte Stix, Matteo Pistillo, Girish Sastry, Marius Hobbhahn, Alejandro Ortega, Mikita Balesni, Annika Hallensleben, Nix Goldowsky-Dill, and Lee Sharkey.
\newblock Ai behind closed doors: a primer on the governance of internal deployment, 2025.
\newblock URL \url{https://arxiv.org/abs/2504.12170}.

\bibitem[Benton et~al.(2024)Benton, Wagner, Christiansen, Anil, Perez, Srivastav, Durmus, Ganguli, Kravec, Shlegeris, Kaplan, Karnofsky, Hubinger, Grosse, Bowman, and Duvenaud]{benton2024sabotageevaluationsfrontiermodels}
Joe Benton, Misha Wagner, Eric Christiansen, Cem Anil, Ethan Perez, Jai Srivastav, Esin Durmus, Deep Ganguli, Shauna Kravec, Buck Shlegeris, Jared Kaplan, Holden Karnofsky, Evan Hubinger, Roger Grosse, Samuel~R. Bowman, and David Duvenaud.
\newblock Sabotage evaluations for frontier models, 2024.
\newblock URL \url{https://arxiv.org/abs/2410.21514}.

\bibitem[Greenblatt et~al.(2024{\natexlab{a}})Greenblatt, Denison, Wright, Roger, MacDiarmid, Marks, Treutlein, Belonax, Chen, Duvenaud, Khan, Michael, Mindermann, Perez, Petrini, Uesato, Kaplan, Shlegeris, Bowman, and Hubinger]{greenblatt2024alignmentfakinglargelanguage}
Ryan Greenblatt, Carson Denison, Benjamin Wright, Fabien Roger, Monte MacDiarmid, Sam Marks, Johannes Treutlein, Tim Belonax, Jack Chen, David Duvenaud, Akbir Khan, Julian Michael, Sören Mindermann, Ethan Perez, Linda Petrini, Jonathan Uesato, Jared Kaplan, Buck Shlegeris, Samuel~R. Bowman, and Evan Hubinger.
\newblock Alignment faking in large language models, 2024{\natexlab{a}}.
\newblock URL \url{https://arxiv.org/abs/2412.14093}.

\bibitem[Meinke et~al.(2025)Meinke, Schoen, Scheurer, Balesni, Shah, and Hobbhahn]{meinke2025frontiermodelscapableincontext}
Alexander Meinke, Bronson Schoen, Jérémy Scheurer, Mikita Balesni, Rusheb Shah, and Marius Hobbhahn.
\newblock Frontier models are capable of in-context scheming, 2025.
\newblock URL \url{https://arxiv.org/abs/2412.04984}.

\bibitem[Eth and Davidson(2025)]{eth2025willairdautomati}
Daniel Eth and Tom Davidson.
\newblock Will {AI} {R\&D} automation cause a software intelligence explosion?, 2025.
\newblock URL \url{https://www.forethought.org/research/will-ai-r-and-d-automation-cause-a-software-intelligence-explosion}.
\newblock Accessed: 2025-04-23.

\bibitem[{Organisation for Economic Co-operation and Development}(2024)]{oecd2024ai}
{Organisation for Economic Co-operation and Development}.
\newblock Assessing potential future artificial intelligence risks, benefits and policy imperatives.
\newblock Technical Report~27, OECD Publishing, November 2024.
\newblock URL \url{https://doi.org/10.1787/3f4e3dfb-en.}
\newblock OECD Artificial Intelligence Papers, No. 27.

\bibitem[Greenblatt et~al.(2024{\natexlab{b}})Greenblatt, Shlegeris, Sachan, and Roger]{greenblatt2024aicontrolimprovingsafety}
Ryan Greenblatt, Buck Shlegeris, Kshitij Sachan, and Fabien Roger.
\newblock {AI} control: Improving safety despite intentional subversion, 2024{\natexlab{b}}.
\newblock URL \url{https://arxiv.org/abs/2312.06942}.

\bibitem[Balesni et~al.(2024)Balesni, Hobbhahn, Lindner, Meinke, Korbak, Clymer, Shlegeris, Scheurer, Stix, Shah, Goldowsky-Dill, Braun, Chughtai, Evans, Kokotajlo, and Bushnaq]{balesni2024evaluationsbasedsafetycasesai}
Mikita Balesni, Marius Hobbhahn, David Lindner, Alexander Meinke, Tomek Korbak, Joshua Clymer, Buck Shlegeris, Jérémy Scheurer, Charlotte Stix, Rusheb Shah, Nicholas Goldowsky-Dill, Dan Braun, Bilal Chughtai, Owain Evans, Daniel Kokotajlo, and Lucius Bushnaq.
\newblock Towards evaluations-based safety cases for {AI} scheming, 2024.
\newblock URL \url{https://arxiv.org/abs/2411.03336}.

\bibitem[Vassilev et~al.(2024)Vassilev, Oprea, Fordyce, and Anderson]{nist2024aml}
Apostol Vassilev, Alina Oprea, Alie Fordyce, and Hyrum Anderson.
\newblock Adversarial machine learning: A taxonomy and terminology of attacks and mitigations.
\newblock Technical Report NIST {AI} 100-2e2023, National Institute of Standards and Technology, January 2024.
\newblock URL \url{https://doi.org/10.6028/NIST.AI.100-2e2023}.
\newblock NIST Trustworthy and Responsible {AI} Report.

\bibitem[Bai et~al.(2022)Bai, Kadavath, Kundu, Askell, Kernion, Jones, Chen, Goldie, Mirhoseini, McKinnon, Chen, Olsson, Olah, Hernandez, Drain, Ganguli, Li, Tran-Johnson, Perez, Kerr, Mueller, Ladish, Landau, Ndousse, Lukosuite, Lovitt, Sellitto, Elhage, Schiefer, Mercado, DasSarma, Lasenby, Larson, Ringer, Johnston, Kravec, Showk, Fort, Lanham, Telleen-Lawton, Conerly, Henighan, Hume, Bowman, Hatfield-Dodds, Mann, Amodei, Joseph, McCandlish, Brown, and Kaplan]{bai2022constitutionalaiharmlessnessai}
Yuntao Bai, Saurav Kadavath, Sandipan Kundu, Amanda Askell, Jackson Kernion, Andy Jones, Anna Chen, Anna Goldie, Azalia Mirhoseini, Cameron McKinnon, Carol Chen, Catherine Olsson, Christopher Olah, Danny Hernandez, Dawn Drain, Deep Ganguli, Dustin Li, Eli Tran-Johnson, Ethan Perez, Jamie Kerr, Jared Mueller, Jeffrey Ladish, Joshua Landau, Kamal Ndousse, Kamile Lukosuite, Liane Lovitt, Michael Sellitto, Nelson Elhage, Nicholas Schiefer, Noemi Mercado, Nova DasSarma, Robert Lasenby, Robin Larson, Sam Ringer, Scott Johnston, Shauna Kravec, Sheer~El Showk, Stanislav Fort, Tamera Lanham, Timothy Telleen-Lawton, Tom Conerly, Tom Henighan, Tristan Hume, Samuel~R. Bowman, Zac Hatfield-Dodds, Ben Mann, Dario Amodei, Nicholas Joseph, Sam McCandlish, Tom Brown, and Jared Kaplan.
\newblock Constitutional {AI}: Harmlessness from {AI} feedback, 2022.
\newblock URL \url{https://arxiv.org/abs/2212.08073}.

\bibitem[Shlegeris(2024{\natexlab{a}})]{shlegeris2024catastrophes}
Buck Shlegeris.
\newblock {AI} catastrophes and rogue deployments, June 2024{\natexlab{a}}.
\newblock URL \url{https://redwoodresearch.substack.com/p/ai-catastrophes-and-rogue-deployments}.
\newblock Redwood Research Blog.

\bibitem[Pan et~al.(2024)Pan, Dai, Fan, and Yang]{pan2024frontieraisystemssurpassed}
Xudong Pan, Jiarun Dai, Yihe Fan, and Min Yang.
\newblock Frontier {AI} systems have surpassed the self-replicating red line, 2024.
\newblock URL \url{https://arxiv.org/abs/2412.12140}.

\bibitem[Knapp and Langill(2015)]{KNAPP201541}
Eric~D. Knapp and Joel~Thomas Langill.
\newblock Chapter 3 - industrial cyber security history and trends.
\newblock In Eric~D. Knapp and Joel~Thomas Langill, editors, \emph{Industrial Network Security (Second Edition)}, pages 41--57. Syngress, Boston, second edition edition, 2015.
\newblock ISBN 978-0-12-420114-9.
\newblock \doi{https://doi.org/10.1016/B978-0-12-420114-9.00003-4}.
\newblock URL \url{https://www.sciencedirect.com/science/article/pii/B9780124201149000034}.

\bibitem[Clymer et~al.(2024{\natexlab{a}})Clymer, Gabrieli, Krueger, and Larsen]{clymer2024safetycasesjustifysafety}
Joshua Clymer, Nick Gabrieli, David Krueger, and Thomas Larsen.
\newblock Safety cases: How to justify the safety of advanced {AI} systems, 2024{\natexlab{a}}.
\newblock URL \url{https://arxiv.org/abs/2403.10462}.

\bibitem[Stein and Dunlop(2024)]{stein2024safe}
Merlin Stein and Connor Dunlop.
\newblock Safe beyond sale: post-deployment monitoring of {AI}, June 2024.
\newblock URL \url{https://www.adalovelaceinstitute.org/blog/post-deployment-monitoring-of-ai/}.
\newblock Ada Lovelace Institute Blog.

\bibitem[O'Brien et~al.(2023)O'Brien, Ee, and Williams]{obrien2023deploymentcorrectionsincidentresponse}
Joe O'Brien, Shaun Ee, and Zoe Williams.
\newblock Deployment corrections: An incident response framework for frontier {AI} models, 2023.
\newblock URL \url{https://arxiv.org/abs/2310.00328}.

\bibitem[Wijk et~al.(2024)Wijk, Lin, Becker, Jawhar, Parikh, Broadley, Chan, Chen, Clymer, Dhyani, Ericheva, Garcia, Goodrich, Jurkovic, Kinniment, Lajko, Nix, Sato, Saunders, Taran, West, and Barnes]{wijk2024rebenchevaluatingfrontierai}
Hjalmar Wijk, Tao Lin, Joel Becker, Sami Jawhar, Neev Parikh, Thomas Broadley, Lawrence Chan, Michael Chen, Josh Clymer, Jai Dhyani, Elena Ericheva, Katharyn Garcia, Brian Goodrich, Nikola Jurkovic, Megan Kinniment, Aron Lajko, Seraphina Nix, Lucas Sato, William Saunders, Maksym Taran, Ben West, and Elizabeth Barnes.
\newblock Re-bench: Evaluating frontier {AI} {R\&D} capabilities of language model agents against human experts, 2024.
\newblock URL \url{https://arxiv.org/abs/2411.15114}.

\bibitem[Owen(2024)]{epoch2024interviewingairesearchersonautomationofairnd}
David Owen.
\newblock Interviewing {AI} researchers on automation of {AI} {R\&D}, 2024.
\newblock URL \url{https://epoch.ai/blog/interviewing-ai-researchers-on-automation-of-ai-rnd}.
\newblock Accessed: 2025-04-16.

\bibitem[Davidson(2023)]{davidson2023compute}
Tom Davidson.
\newblock What a compute-centric framework says about takeoff speeds.
\newblock Technical report, Open Philanthropy, June 2023.
\newblock URL \url{https://www.openphilanthropy.org/research/what-a-compute-centric-framework-says-about-takeoff-speeds/}.
\newblock Research Report.

\bibitem[DeepSeek-{AI} et~al.(2025)DeepSeek-{AI}, Liu, Feng, Xue, Wang, Wu, Lu, Zhao, Deng, Zhang, Ruan, Dai, Guo, Yang, Chen, Ji, Li, Lin, Dai, Luo, Hao, Chen, Li, Zhang, Bao, Xu, Wang, Zhang, Ding, Xin, Gao, Li, Qu, Cai, Liang, Guo, Ni, Li, Wang, Chen, Chen, Yuan, Qiu, Li, Song, Dong, Hu, Gao, Guan, Huang, Yu, Wang, Zhang, Xu, Xia, Zhao, Wang, Zhang, Li, Wang, Zhang, Zhang, Tang, Li, Tian, Huang, Wang, Zhang, Wang, Zhu, Chen, Du, Chen, Jin, Ge, Zhang, Pan, Wang, Xu, Zhang, Chen, Li, Lu, Zhou, Chen, Wu, Ye, Ye, Ma, Wang, Zhou, Yu, Zhou, Pan, Wang, Yun, Pei, Sun, Xiao, Zeng, Zhao, An, Liu, Liang, Gao, Yu, Zhang, Li, Jin, Wang, Bi, Liu, Wang, Shen, Chen, Zhang, Chen, Nie, Sun, Wang, Cheng, Liu, Xie, Liu, Yu, Song, Shan, Zhou, Yang, Li, Su, Lin, Li, Wang, Wei, Zhu, Zhang, Xu, Xu, Huang, Li, Zhao, Sun, Li, Wang, Yu, Zheng, Zhang, Shi, Xiong, He, Tang, Piao, Wang, Tan, Ma, Liu, Guo, Wu, Ou, Zhu, Wang, Gong, Zou, He, Zha, Xiong, Ma, Yan, Luo, You, Liu, Zhou, Wu, Ren, Ren, Sha, Fu, Xu, Huang, Zhang, Xie, Zhang,
  Hao, Gou, Ma, Yan, Shao, Xu, Wu, Zhang, Li, Gu, Zhu, Liu, Li, Xie, Song, Gao, and Pan]{deepseekai2025deepseekv3technicalreport}
DeepSeek-{AI}, Aixin Liu, Bei Feng, Bing Xue, Bingxuan Wang, Bochao Wu, Chengda Lu, Chenggang Zhao, Chengqi Deng, Chenyu Zhang, Chong Ruan, Damai Dai, Daya Guo, Dejian Yang, Deli Chen, Dongjie Ji, Erhang Li, Fangyun Lin, Fucong Dai, Fuli Luo, Guangbo Hao, Guanting Chen, Guowei Li, H.~Zhang, Han Bao, Hanwei Xu, Haocheng Wang, Haowei Zhang, Honghui Ding, Huajian Xin, Huazuo Gao, Hui Li, Hui Qu, J.~L. Cai, Jian Liang, Jianzhong Guo, Jiaqi Ni, Jiashi Li, Jiawei Wang, Jin Chen, Jingchang Chen, Jingyang Yuan, Junjie Qiu, Junlong Li, Junxiao Song, Kai Dong, Kai Hu, Kaige Gao, Kang Guan, Kexin Huang, Kuai Yu, Lean Wang, Lecong Zhang, Lei Xu, Leyi Xia, Liang Zhao, Litong Wang, Liyue Zhang, Meng Li, Miaojun Wang, Mingchuan Zhang, Minghua Zhang, Minghui Tang, Mingming Li, Ning Tian, Panpan Huang, Peiyi Wang, Peng Zhang, Qiancheng Wang, Qihao Zhu, Qinyu Chen, Qiushi Du, R.~J. Chen, R.~L. Jin, Ruiqi Ge, Ruisong Zhang, Ruizhe Pan, Runji Wang, Runxin Xu, Ruoyu Zhang, Ruyi Chen, S.~S. Li, Shanghao Lu, Shangyan Zhou,
  Shanhuang Chen, Shaoqing Wu, Shengfeng Ye, Shengfeng Ye, Shirong Ma, Shiyu Wang, Shuang Zhou, Shuiping Yu, Shunfeng Zhou, Shuting Pan, T.~Wang, Tao Yun, Tian Pei, Tianyu Sun, W.~L. Xiao, Wangding Zeng, Wanjia Zhao, Wei An, Wen Liu, Wenfeng Liang, Wenjun Gao, Wenqin Yu, Wentao Zhang, X.~Q. Li, Xiangyue Jin, Xianzu Wang, Xiao Bi, Xiaodong Liu, Xiaohan Wang, Xiaojin Shen, Xiaokang Chen, Xiaokang Zhang, Xiaosha Chen, Xiaotao Nie, Xiaowen Sun, Xiaoxiang Wang, Xin Cheng, Xin Liu, Xin Xie, Xingchao Liu, Xingkai Yu, Xinnan Song, Xinxia Shan, Xinyi Zhou, Xinyu Yang, Xinyuan Li, Xuecheng Su, Xuheng Lin, Y.~K. Li, Y.~Q. Wang, Y.~X. Wei, Y.~X. Zhu, Yang Zhang, Yanhong Xu, Yanhong Xu, Yanping Huang, Yao Li, Yao Zhao, Yaofeng Sun, Yaohui Li, Yaohui Wang, Yi~Yu, Yi~Zheng, Yichao Zhang, Yifan Shi, Yiliang Xiong, Ying He, Ying Tang, Yishi Piao, Yisong Wang, Yixuan Tan, Yiyang Ma, Yiyuan Liu, Yongqiang Guo, Yu~Wu, Yuan Ou, Yuchen Zhu, Yuduan Wang, Yue Gong, Yuheng Zou, Yujia He, Yukun Zha, Yunfan Xiong, Yunxian Ma, Yuting
  Yan, Yuxiang Luo, Yuxiang You, Yuxuan Liu, Yuyang Zhou, Z.~F. Wu, Z.~Z. Ren, Zehui Ren, Zhangli Sha, Zhe Fu, Zhean Xu, Zhen Huang, Zhen Zhang, Zhenda Xie, Zhengyan Zhang, Zhewen Hao, Zhibin Gou, Zhicheng Ma, Zhigang Yan, Zhihong Shao, Zhipeng Xu, Zhiyu Wu, Zhongyu Zhang, Zhuoshu Li, Zihui Gu, Zijia Zhu, Zijun Liu, Zilin Li, Ziwei Xie, Ziyang Song, Ziyi Gao, and Zizheng Pan.
\newblock Deepseek-v3 technical report, 2025.
\newblock URL \url{https://arxiv.org/abs/2412.19437}.

\bibitem[{AlphaProof and AlphaGeometry teams}(2024)]{deepmind2024imo}
{AlphaProof and AlphaGeometry teams}.
\newblock {AI} achieves silver-medal standard solving international mathematical olympiad problems, July 2024.
\newblock URL \url{https://deepmind.google/discover/blog/ai-solves-imo-problems-at-silver-medal-level/}.
\newblock DeepMind Blog.

\bibitem[Pilz et~al.(2024)Pilz, Heim, and Brown]{pilz2024increasedcomputeefficiencydiffusion}
Konstantin Pilz, Lennart Heim, and Nicholas Brown.
\newblock Increased compute efficiency and the diffusion of {AI} capabilities, 2024.
\newblock URL \url{https://arxiv.org/abs/2311.15377}.

\bibitem[Armstrong et~al.(2013)Armstrong, Bostrom, and Shulman]{armstrong2013racing}
Stuart Armstrong, Nick Bostrom, and Carl Shulman.
\newblock Racing to the precipice: A model of artificial intelligence development.
\newblock Technical Report 2013-1, Future of Humanity Institute, University of Oxford, 2013.

\bibitem[Clymer et~al.(2024{\natexlab{b}})Clymer, Wijk, and Barnes]{the-rogue-replication-threat-model}
Josh Clymer, Hjalmar Wijk, and Beth Barnes.
\newblock The rogue replication threat model.
\newblock \url{https://metr.org/blog/2024-11-12-rogue-replication-threat-model/}, 11 2024{\natexlab{b}}.

\bibitem[{Anthropic}(2024)]{anthropic2024rspupdate}
{Anthropic}.
\newblock Announcing our updated responsible scaling policy, October 2024.
\newblock URL \url{https://www.anthropic.com/news/announcing-our-updated-responsible-scaling-policy}.
\newblock Anthropic Blog.

\bibitem[Eiras et~al.(2024)Eiras, Petrov, Vidgen, Schroeder, Pizzati, Elkins, Mukhopadhyay, Bibi, Purewal, Botos, Steibel, Keshtkar, Barez, Smith, Guadagni, Chun, Cabot, Imperial, Nolazco, Landay, Jackson, Torr, Darrell, Lee, and Foerster]{eiras2024risksopportunitiesopensourcegenerative}
Francisco Eiras, Aleksandar Petrov, Bertie Vidgen, Christian Schroeder, Fabio Pizzati, Katherine Elkins, Supratik Mukhopadhyay, Adel Bibi, Aaron Purewal, Csaba Botos, Fabro Steibel, Fazel Keshtkar, Fazl Barez, Genevieve Smith, Gianluca Guadagni, Jon Chun, Jordi Cabot, Joseph Imperial, Juan~Arturo Nolazco, Lori Landay, Matthew Jackson, Phillip H.~S. Torr, Trevor Darrell, Yong Lee, and Jakob Foerster.
\newblock Risks and opportunities of open-source generative {AI}, 2024.
\newblock URL \url{https://arxiv.org/abs/2405.08597}.

\bibitem[Kak and West(2023)]{kak2023landscape}
Amba Kak and Sarah~Myers West.
\newblock {AI} now 2023 landscape: Confronting tech power.
\newblock Technical report, {AI} Now Institute, April 2023.
\newblock URL \url{https://ainowinstitute.org/2023-landscape}.
\newblock Technical Report.

\bibitem[Seger et~al.(2023)Seger, Dreksler, Moulange, Dardaman, Schuett, Wei, Winter, Arnold, hÉigeartaigh, Korinek, Anderljung, Bucknall, Chan, Stafford, Koessler, Ovadya, Garfinkel, Bluemke, Aird, Levermore, Hazell, and Gupta]{seger2023opensourcinghighlycapablefoundation}
Elizabeth Seger, Noemi Dreksler, Richard Moulange, Emily Dardaman, Jonas Schuett, K.~Wei, Christoph Winter, Mackenzie Arnold, Seán~Ó hÉigeartaigh, Anton Korinek, Markus Anderljung, Ben Bucknall, Alan Chan, Eoghan Stafford, Leonie Koessler, Aviv Ovadya, Ben Garfinkel, Emma Bluemke, Michael Aird, Patrick Levermore, Julian Hazell, and Abhishek Gupta.
\newblock Open-sourcing highly capable foundation models: An evaluation of risks, benefits, and alternative methods for pursuing open-source objectives, 2023.
\newblock URL \url{https://arxiv.org/abs/2311.09227}.

\bibitem[Corsi et~al.(2024)Corsi, Kilian, and Mallah]{corsi2024considerationsinfluencingoffensedefensedynamics}
Giulio Corsi, Kyle Kilian, and Richard Mallah.
\newblock Considerations influencing offense-defense dynamics from artificial intelligence, 2024.
\newblock URL \url{https://arxiv.org/abs/2412.04029}.

\bibitem[Bernardi et~al.(2025)Bernardi, Mukobi, Greaves, Heim, and Anderljung]{bernardi2025societaladaptationadvancedai}
Jamie Bernardi, Gabriel Mukobi, Hilary Greaves, Lennart Heim, and Markus Anderljung.
\newblock Societal adaptation to advanced {AI}, 2025.
\newblock URL \url{https://arxiv.org/abs/2405.10295}.

\bibitem[Wasil et~al.(2024{\natexlab{a}})Wasil, Berglund, Reed, Plueckebaum, and Smith]{wasil2024understanding}
Akash Wasil, Lukas Berglund, Tom Reed, Miro Plueckebaum, and Everett Smith.
\newblock Understanding frontier {AI} capabilities and risks through semi-structured interviews.
\newblock Technical report, SSRN, July 2024{\natexlab{a}}.
\newblock URL \url{https://ssrn.com/abstract=4881729}.
\newblock SSRN Working Paper.

\bibitem[Wasil et~al.(2024{\natexlab{b}})Wasil, Smith, Katzke, and Bullock]{wasil2024aiemergencypreparednessexamining}
Akash Wasil, Everett Smith, Corin Katzke, and Justin Bullock.
\newblock {AI} emergency preparedness: Examining the federal government's ability to detect and respond to ai-related national security threats, 2024{\natexlab{b}}.
\newblock URL \url{https://arxiv.org/abs/2407.17347}.

\bibitem[Nevo et~al.(2024)Nevo, Lahav, Karpur, Bar-On, Bradley, and Alstott]{RR-A2849-1}
Sella Nevo, Dan Lahav, Ajay Karpur, Yogev Bar-On, Henry~Alexander Bradley, and Jeff Alstott.
\newblock \emph{Securing {AI} Model Weights: Preventing Theft and Misuse of Frontier Models}.
\newblock RAND Corporation, Santa Monica, CA, 2024.
\newblock \doi{10.7249/RRA2849-1}.

\bibitem[Shlegeris(2024{\natexlab{b}})]{shlegeris2024access}
Buck Shlegeris.
\newblock Access to powerful {AI} might make computer security radically easier, June 2024{\natexlab{b}}.
\newblock URL \url{https://redwoodresearch.substack.com/p/access-to-powerful-ai-might-make}.
\newblock Redwood Research Blog.

\bibitem[Davidson et~al.(2023)Davidson, Denain, Villalobos, and Bas]{davidson2023aicapabilitiessignificantlyimproved}
Tom Davidson, Jean-Stanislas Denain, Pablo Villalobos, and Guillem Bas.
\newblock {AI} capabilities can be significantly improved without expensive retraining, 2023.
\newblock URL \url{https://arxiv.org/abs/2312.07413}.

\bibitem[{Epoch AI}(2025)]{epoch2025benchmarking}
{Epoch AI}.
\newblock {AI} benchmarking hub, April 2025.
\newblock URL \url{https://epoch.ai/data/ai-benchmarking-dashboard}.
\newblock Last updated April 16, 2025.

\end{thebibliography}
 
\end{document}